\pdfoutput=1
\PassOptionsToPackage{table,dvipsnames}{xcolor}

\documentclass[11pt]{article}

\usepackage[preprint]{acl}

\usepackage{times}
\usepackage{latexsym}

\usepackage[T1]{fontenc}

\usepackage[utf8]{inputenc}

\usepackage{microtype}

\usepackage{inconsolata}

\usepackage{graphicx}
\usepackage{multirow}
\usepackage{booktabs}
\usepackage{amsmath}
\DeclareMathOperator*{\softmax}{softmax} 
\usepackage{xspace}
\usepackage{pifont}
\usepackage{soul}   
\usepackage[normalem]{ulem}   
\usepackage{booktabs}
\usepackage[most]{tcolorbox}
\usepackage{arydshln}
\newcommand{\nrr}{StandardRR\xspace}     
\newcommand{\rr}{ReasonRR\xspace}    
\newcommand{\rrnr}{ReasonRR-NoReason\xspace} 

\title{Don’t ``Overthink'' Passage Reranking: Is Reasoning Truly Necessary?}

\author{
 \textbf{Nour Jedidi\textsuperscript{1}} \quad
 \textbf{Yung-Sung Chuang\textsuperscript{2}} \quad
 \textbf{James Glass\textsuperscript{2}} \quad 
 \textbf{Jimmy Lin\textsuperscript{3}} \quad 
\\
 \textsuperscript{1}MIT Lincoln Laboratory  \quad 
  \textsuperscript{2}Massachusetts Institute of Technology \quad 
  \textsuperscript{3}University of Waterloo \quad 
\\
\texttt{nour.jedidi@ll.mit.edu} \quad
}

\begin{document}
\maketitle

\begin{abstract}
With the growing success of reasoning models across complex natural language tasks, researchers in the Information Retrieval (IR) community have begun exploring how similar reasoning capabilities can be integrated into passage rerankers built on Large Language Models (LLMs). These methods typically employ an LLM to produce an explicit, step-by-step reasoning process before arriving at a final relevance prediction. But, \textit{does reasoning actually improve reranking accuracy?} In this paper, we dive deeper into this question, studying the impact of the reasoning process by comparing reasoning-based pointwise rerankers (\rr) to standard, non-reasoning pointwise rerankers (\nrr) under identical training conditions, and observe that \nrr generally outperforms \rr. Building on this observation, we then study the importance of reasoning to \rr by disabling its reasoning process (\rrnr), and find that \rrnr is surprisingly more effective than \rr. Examining the cause of this result, our findings reveal that reasoning-based rerankers are limited by the LLM's reasoning process, which pushes it toward polarized relevance scores and thus fails to consider the \textit{partial} relevance of passages, a key factor for the accuracy of pointwise rerankers.

\end{abstract}

\section{Introduction}
\begin{figure}[t]
\centering
\includegraphics[width=1.0\linewidth]{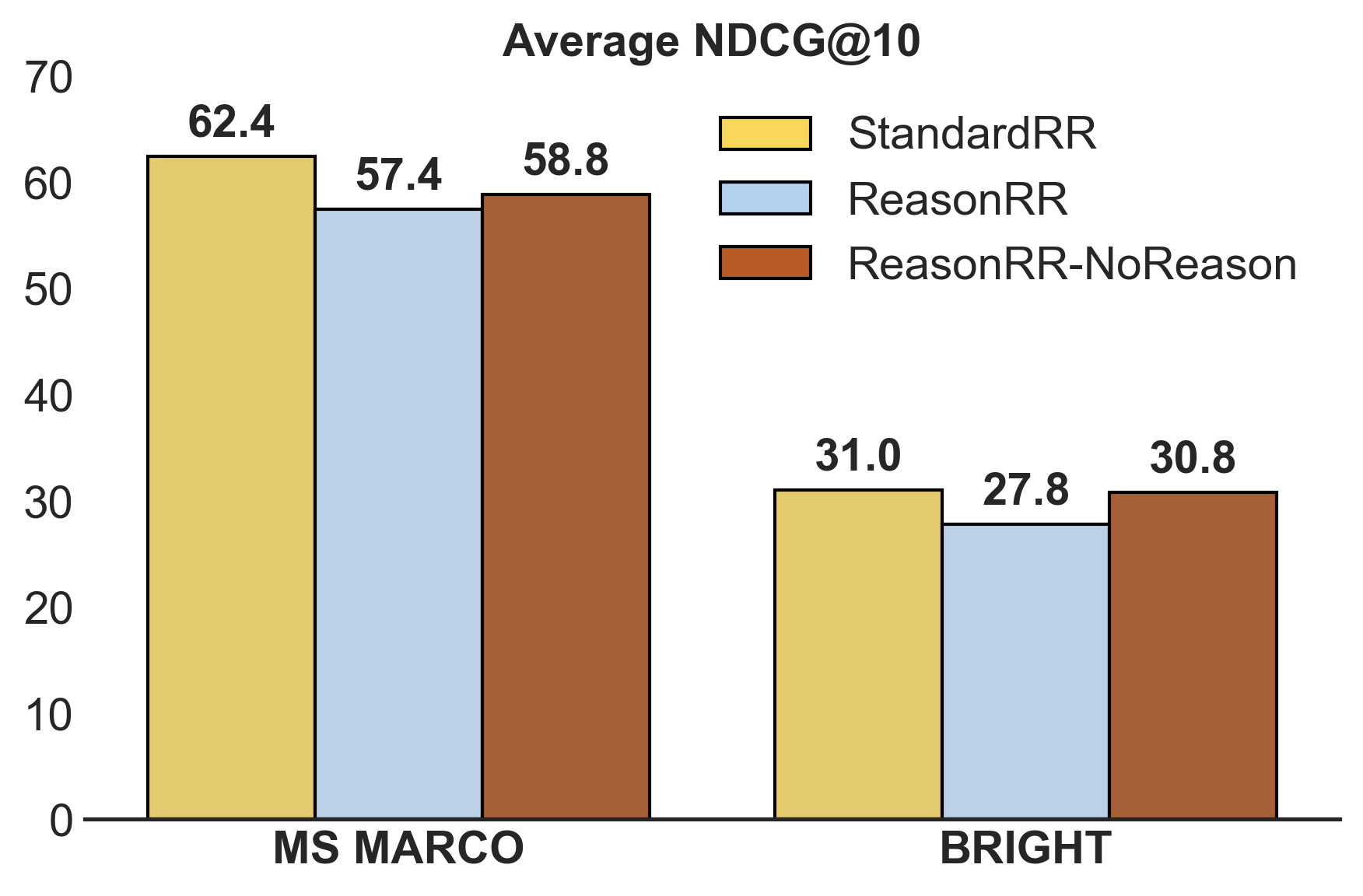}
\caption{Average NDCG@10 of reasoning pointwise rerankers (\rr) compared to their non-reasoning variants (\nrr and \rrnr) on MS MARCO and BRIGHT.}
\label{fig:teaser}
\end{figure}
Recently, there has been a surge of interest in reasoning models such as DeepSeek-R1~\cite{guo2025deepseek}, OpenAI's o3, Qwen3~\cite{yang2025qwen3}, and others. By generating an explicit reasoning process — i.e., a chain-of-thought (CoT) —  prior to producing its final response, reasoning models have shown strong performance across a wide range of complex natural language tasks such as mathematics~\citep{yang2024qwen2}.

Following the success of reasoning models, researchers in the Information Retrieval (IR) community have begun to explore how incorporating a reasoning process can improve Large Language Model (LLM) based retrieval systems~\cite{yan2025o1, ji2025learning, shao2025reasonir, weller2025rank1, zhuang2025rank}, especially with the introduction of reasoning-intensive retrieval benchmarks such as BRIGHT~\cite{su2025bright}. In particular, recent work has explored incorporating a reasoning process to improve LLM-based pointwise~\cite{weller2025rank1} and setwise~\cite{zhuang2025rank} rerankers, showing promising results on reasoning-intensive retrieval tasks. 

However, the importance of the explicit reasoning processes in rerankers when compared to standard, non-reasoning rerankers under \emph{identical} training conditions remains an open question. In this paper, we study the necessity of this explicit reasoning processes and ask: \textit{Does scaling test-time compute — via generation of reasoning tokens prior to making a relevance prediction — actually improve reranking accuracy?} To investigate this, we examine two perspectives: 

\begin{itemize}
  \setlength\itemsep{0em}
  \item  \textbf{RQ1}:  Under identical training setups (e.g., training data and backbone LLM), how do reasoning rerankers compare against standard, non-reasoning rerankers?
  \item \textbf{RQ2}: How does disabling the reasoning process of a reasoning reranker affect its reranking accuracy?
\end{itemize}

To answer these research questions, we train and evaluate three  reranker variants: (1) \textbf{\nrr}, a standard LLM-based pointwise reranker that directly classifies query-passage pairs as relevant or non-relevant~\cite{nogueira-etal-2020-document, ma2024fine}; (2) \textbf{\rr}, our reproduction of Rank1 \cite{weller2025rank1}, which  generates a reasoning chain prior to making a relevance prediction; and (3) \textbf{\rrnr}, a modified version of \rr, in which the explicit reasoning process is disabled at inference time by providing a forced reasoning process, effectively transforming \rr into a standard pointwise reranker. The central findings of our experiments can be summarized as follows and are shown in Figure \ref{fig:teaser}:

\begin{itemize}
  \setlength\itemsep{0em}
  \item   Under identical training setups, we find no general advantage of the reasoning process for pointwise reranking. While certain domains and LLM scales benefit from reasoning, on average, \nrr did better on in-domain and out-of-domain datasets versus \rr.
  \item In fact, we find that reasoning can even degrade effectiveness for rerankers explicitly trained to reason (i.e., \rr). As shown in Figure \ref{fig:teaser}, \rrnr outperforms \rr by 1.4 points in-domain (MS MARCO) and 3 points out-of-domain (BRIGHT) in terms of NDCG@10.
  \item  Further investigation suggests that this likely stems from the reasoning process forcing the model towards \textit{polarized} relevance scores which does not account for the \textit{partial} relevance of passages. Our results show that while \rr is a better relevance classifier than \rrnr, \rrnr placed more emphasis on partial relevance scores, contributing to its better reranking accuracy.
  \item   While we find that self-consistency \cite{wang2023selfconsistency} can bring about improvements to \rr, it is still outperformed by \nrr. Given this, our results suggest that practitioners are better served by simpler methods like \nrr, which is more accurate and cost-effective.
\end{itemize}

Our findings build upon recent research in the NLP community which question the necessity of the reasoning process for LLMs  \cite{ma2025reasoning}. We hope our work not only encourages future work that can improve reasoning for reranking tasks, but also highlights the importance of comparing against strong, simple baselines when developing new methodologies.

\section{Background}

In this section, we provide brief background on pointwise rerankers (\nrr) and reasoning pointwise rerankers (\rr), as they form the basis of our study. We emphasize that \rr is our reproduction of Rank1 \cite{weller2025rank1}; however, we refer to it as \rr to maintain consistency with the usage of \nrr throughout the paper.

\paragraph{Preliminaries.} The goal of information retrieval (IR) is to identify relevant passages from a large collection of $n$ texts, denoted by $\mathcal{C}=\{P_1, P_2, \ldots, P_n\}$, given a user-issued query, $q$. Current IR systems typically employ a multi-stage pipeline where a first-stage retriever fetches an initial set of $k$ passages from $\mathcal{C}$ and a \textit{reranker} reorders the top-$k$ passages $\{P_1, P_2, \ldots, P_k\}$, where $k \ll n$, to produce a more accurate ranking.

\paragraph{\nrr.} \nrr is trained as a pointwise reranker, independently producing a relevance score, $R$, for a given query-passage pair. To train \nrr, the simplest approach is to directly fine-tune an LLM to produce the tokens ``true'' or ``false'' given a dataset of (query, passage, relevance label) triples, where  ``true'' and ``false'' denote relevant or not relevant, respectively.

At inference, for each query-passage pair $(q, P_{i})$ in the top-$k$, the probability of relevance, $R$, is computed by applying a softmax exclusively to the logits corresponding to the tokens ``true'' and ``false'':

\begin{equation}
\begin{aligned}
R = \softmax\big(&\text{z}_{\text{true}}(q, P_{i}), \text{z}_{\text{false}}(q, P_{i})\big)_{\text{true}}
\end{aligned}
\label{eq: simple_pr}
\end{equation}

\noindent
Here, $\text{z}_{\text{true}}(q, P_{i})$ and $\text{z}_{\text{false}}(q, P_{i})$ denote the logits assigned by the LLM for the ``true'' and ``false'' tokens,  given input $(q, P_{i})$. The subscript ``true'' after the softmax normalization indicates that only the probability assigned to the token ``true'' is considered for $R$. The passages are then sorted in descending order of $R$. We note that recent methods, such as RankLLaMA~\cite{ma2024fine}, train pointwise rerankers using hard negatives sampled from the top-ranking results of a first-stage retriever. However, as our goal is to keep the training setup identical to that of \rr, which we describe next, we do not consider hard negatives.

\paragraph{\rr.} \rr builds upon the setup described for \nrr by fine-tuning an LLM to first generate a reasoning process, $r$, before  producing the tokens ``true'' or ``false''. To do so, \rr is fine-tuned with a dataset of (query, passage, $r$, relevance label) quadruples.

Following Equation~\ref{eq: simple_pr}, $R$ is again computed by considering the softmax over the logits of the ``true'' and ``false'' tokens, but in this case, $R$ also considers the LLM's generated reasoning process, $r$:

\begin{equation}
\begin{aligned}
R = \softmax\big(&\text{z}_{\text{true}}(q, P_{i}, r_{i}), \\
                         &\text{z}_{\text{false}}(q, P_{i}, r_{i})\big)_{\text{true}}
\end{aligned}
\end{equation}

\noindent
where $r_{i}$ is the reasoning process generated for input $(q, P_{i})$. The passages are then reordered as described for \nrr.
\section{Does Reasoning Improve Rerankers?}

In this section, we study the impact of reasoning on pointwise rerankers through two different lenses: (1) how does \nrr compare to \rr when trained under the same settings? And, (2) how is \rr's reranking accuracy affected if we forcefully remove its reasoning process (\rrnr)? Through these two perspectives, we hope to shed light on different ways reasoning may influence reranking accuracy.

\subsection{RQ1: \nrr vs. \rr}
\label{sec: rq1}

\begin{table}[t!]
\centering
\resizebox{\columnwidth}{!}{%
\begin{tabular}{l|cc|ccc}
\toprule
& \multicolumn{2}{c|}{MS MARCO v1} & \multicolumn{3}{c}{MS MARCO v2} \\
\cmidrule(r){2-3} \cmidrule(r){4-6}
& DL19 & DL20 & DL21 & DL22 & DL23 \\
\midrule
BM25 & 50.6 & 48.0 & 44.6 & 26.9 & 26.3 \\
\midrule
+ Qwen2.5-1.5B &&&&& \\
\quad \nrr & \textbf{73.1}	 & \textbf{69.4}  &	\textbf{68.9}  &	\textbf{50.7}	 & \textbf{44.2} \\
\quad \rr & 68.7	& 63.1	& 65.7	& 43.3	& 38.8 \\
\midrule
+ Qwen2.5-3B  &&&&& \\
\quad \nrr & \textbf{72.5}	& \textbf{68.9}	& \textbf{69.4}	& \textbf{51.4}	& \textbf{45.5} \\
\quad \rr& 70.4 & 66.4 & 65.9 & 45.2 & 41.3\\
\midrule
+ Qwen2.5-7B &&&&& \\
\quad \nrr & \textbf{74.6} & \textbf{70.0} & \textbf{70.9} & \textbf{50.3} & \textbf{46.3} \\
\quad \rr & 70.3 & 64.3 & 65.9 & 45.6 & 41.1\\
\bottomrule
\end{tabular}
}
\caption{In-domain performance of \nrr versus \rr. Each Qwen2.5 model reranks the top-100 passages from BM25.}
\label{tab:msmarco}
\end{table}

\begin{table*}[t!]
\centering
\resizebox{\textwidth}{!}{
\begin{tabular}{l|rrrrrrr|rr|rrr|r}
\toprule
& \multicolumn{7}{c|}{StackExchange} & \multicolumn{2}{c|}{Coding} & \multicolumn{3}{c|}{Theorem-based} & \multirow{2}{*}{Avg.} \\
\cmidrule(r){2-8} \cmidrule(r){9-10} \cmidrule(r){11-13}
& Bio. & Earth. & Econ. & Psy. & Rob. & Stack. & Sus. & Leet. & Pony & AoPS & TheoQ. & TheoT. & \\
\midrule
BM25 + GPT-4 CoT              & 53.6 &	54.1 &	24.3 &	38.7 &	18.9 &	27.7 &	26.3 &	19.3 &	17.6 &	3.9 &	19.2 &	20.8 &	27.0 \\
\midrule
+ Qwen2.5-1.5B &&&&&&&&&&&&&  \\
\quad \nrr   & \textbf{37.0} &	\textbf{21.7} &	\textbf{16.8} &	23.1 &	\textbf{16.1} &	10.0 &	\textbf{26.3} &	2.6 &	\textbf{30.6} &	1.8 &	\textbf{16.1} &	\textbf{26.1} &	\textbf{19.0}	 \\ 
\quad \rr     & 32.5	& 20.3 & 	12.3 & 	\textbf{25.5} & 	11.1 & 	\textbf{15.3} & 	23.5 & 	\textbf{6.6} & 	12.3 & 	\textbf{3.4} & 	10.6 & 	13.7 &	15.6	 \\
\midrule
+ Qwen2.5-3B &&&&&&&&&&&&&   \\
\quad \nrr    & \textbf{41.6} &	27.1 &	  \textbf{20.9} &	31.9 &	\textbf{22.2} &	16.9 &	\textbf{30.3} &	\textbf{13.2} &	\textbf{42.0} &	2.7 &	16.2 &	30.6 &	\textbf{24.6}	 \\
\quad \rr     &  37.3 &	\textbf{27.8} &	20.7 &	\textbf{33.1} &	18.3 &	\textbf{24.3} &	25.2 &	11.3 &	26.2 &	\textbf{4.7} &	\textbf{20.7} &	\textbf{34.0} &	23.6\\
\midrule
+ Qwen2.5-7B &&&&&&&&&&&&&   \\
\quad \nrr          &  \textbf{47.1} &	\textbf{38.0} &	\textbf{28.1} &	\textbf{44.1} &	\textbf{26.1} &	\textbf{29.5} &	\textbf{36.5} &	\textbf{19.3} &	\textbf{37.5} &	4.6 &	\textbf{22.4} &	\textbf{39.4} &	\textbf{31.0}	 \\ 
\quad \rr            & 47.0 &	35.4 &	24.0 &	35.2 &	20.0 &	25.2 &	31.0 & 	15.1&	36.0 &	\textbf{5.9}	 & 22.2	& 36.6 &	27.8\\
\bottomrule
\end{tabular}
}
\caption{Out-of-domain performance of \nrr versus \rr. Each Qwen2.5 model reranks the top-100 passages from BM25 + GPT-4 CoT.}
\label{tab:bright}
\end{table*}

Our first experiment aims to understand the importance of reasoning from the training perspective. Specifically, if we train \nrr on the exact same data as \rr, but omit the reasoning chain, how does performance compare? To answer this research question, we train pointwise rerankers of varying sizes, with and without reasoning chains.

\paragraph{Experiment Setup.} To train the rerankers, we leverage the training data provided by \citet{weller2025rank1}.\footnote{\url{https://huggingface.co/datasets/jhu-clsp/rank1-training-data}} The dataset augments MS MARCO \cite{bajaj2016ms} with reasoning chains generated by Deepseek R1 \cite{guo2025deepseek}, which include R1's final relevance predictions. The dataset consists of approximately 386K quadruples in the following format: (query, passage, R1's reasoning chain, relevance label). 

For the backbone LLM, we leverage the Qwen2.5 base models \cite{yang2024qwen2} ranging from 1.5B to 7B model sizes. To train \rr we fine-tune using LoRA \cite{hu2022lora} for one epoch with rank 32 and alpha 64. To train \nrr we follow the same setup, but only use the (query, passage, relevance label) triples, omitting R1's reasoning chain. 

We evaluate \nrr  and \rr on in-domain and out-of-domain retrieval datasets. For in-domain evaluation, we leverage passage ranking datasets based on MS MARCO v1—TREC DL19 and TREC DL20 \cite{Craswell2020OverviewOT, Craswell2021OverviewOT}—and based on MS MARCO v2—TREC DL21, TREC DL22, and TREC DL23 \cite{Craswell2021OverviewOT_21, Craswell2022OverviewOT, Craswell2023OverviewOT}. For out-of-domain evaluation, we focus on BRIGHT \cite{su2025bright}, a reasoning-intensive retrieval benchmark. We report NDCG@10, the official metric for both the MS MARCO and BRIGHT datasets.

At inference, models rerank the top-100 passages retrieved by BM25. For BRIGHT, models rerank passages retrieved by BM25 using queries expanded with GPT-4 CoT; however, following~\citet{weller2025rank1}, the rerankers are \textit{not} provided the GPT-4 CoT. For MS MARCO, we implement BM25 using Pyserini \cite{lin2021pyserini} and for BRIGHT, we follow the implementation from the BRIGHT codebase. LLM training was performed using HuggingFace \cite{wolf2019huggingface} and inference with vLLM \cite{kwon2023efficient}.

\paragraph{Results.} In Tables \ref{tab:msmarco} and \ref{tab:bright} we present the evaluation results for both in-domain and out-of-domain retrieval tasks. As shown in Appendix \ref{sec:reproduction_rank1}, \rr is comparable to Rank1, achieving a similar NDCG@10 (27.8 versus 27.5, respectively), confirming our implementation is valid.

On MS MARCO, we find that \nrr outperforms  \rr by an average of 5.3, 3.7, and 5 points across the 1.5B, 3B, and 7B model sizes, respectively. Surprisingly, on BRIGHT, we find a similar story: \nrr outperforms \rr, achieving 3.4, 1, and 3.2 points higher average NDCG@10 across the 1.5B, 3B, and 7B model sizes.  However, while in-domain \nrr always outperformed \rr, out-of-domain, the results suggest that reasoning \emph{can} be beneficial depending on the model scale and domain. For example, at the smaller model scales (1.5B and 3B), \rr achieves higher NDCG@10 versus \nrr on the Psychology (Psy.), Stack Overflow (Stack.), and  AoPS datasets. At the 7B scale, while \nrr begins to consistently outperform \rr, we find that \rr still performs better on the AoPs dataset. 

All in all, these results suggest that while reasoning can improve rerankers for certain model sizes and domains, training a \rr-style pointwise reranker does not provide any general advantage versus \nrr. 

\subsection{RQ2: How Important is the Reasoning Process to \rr?}
\label{sec:rq2}


\begin{table}[t!]
\centering
\resizebox{\columnwidth}{!}{%
\begin{tabular}{c|l|cc}
\toprule
Qwen2.5 & Method & MS MARCO & BRIGHT \\
\midrule
\multirow{3}{*}{1.5B} & \nrr& 61.3 & 19.0 \\
\cdashline{2-4}
& \rr & 55.9 & \textbf{15.6} \\
& \rrnr & \textbf{56.7} & 11.6 \\
\midrule
\multirow{3}{*}{3B} & \nrr & 61.5 & 24.6\\
\cdashline{2-4}
& \rr  & 57.8 & \textbf{23.6}\\
& \rrnr & \textbf{58.3} & 23.4 \\
\midrule
\multirow{3}{*}{7B} & \nrr & 62.4 & 31.0 \\
\cdashline{2-4}
& \rr & 57.4 & 27.8 \\
& \rrnr   & \textbf{58.8} & \textbf{30.8} \\
\bottomrule
\end{tabular}
}
\caption{Studying the effect of removing the reasoning process from pointwise rerankers with reasoning. Results on MS MARCO and BRIGHT represent an average across the corresponding datasets. \textbf{Bold} results denote best between \rr and \rrnr. See Appendix \ref{sec:full_results_ablations} for results on individual datasets.}
\label{tab:no_reasoning}
\end{table}

\begin{table*}[t]
\centering
\resizebox{\textwidth}{!}{
\begin{tabular}{l|ccc|ccc|ccc|ccc|ccc}
\toprule
& \multicolumn{6}{c|}{MS MARCO v1} & \multicolumn{9}{c}{MS MARCO v2} \\
\cmidrule(r){2-7} \cmidrule(r){8-16}
& \multicolumn{3}{c|}{DL19} & \multicolumn{3}{c|}{DL20} & \multicolumn{3}{c|}{DL21} & \multicolumn{3}{c|}{DL22} & \multicolumn{3}{c}{DL23} \\
\cmidrule(r){2-16}
& P & R & F1 & P & R & F1 & P & R & F1 & P & R & F1 & P & R & F1 \\
\midrule
\nrr   & \textbf{71.4} & 80.3 & \textbf{75.6} & \textbf{54.5} & 79.3 & \textbf{64.6} & \textbf{56.4} & 87.1 & 
\textbf{68.5} & \textbf{49.1} & 70.0 & \textbf{57.6} & \textbf{39.7} & 66.9 & \textbf{49.8} \\
\midrule
\rr    & 65.9 & 82.4 & 73.2 & 49.2 & 82.1 & 61.5 & 54.0 & 89.2 & 67.3 & 41.6 & 73.2 & 53.1 & 35.6 & 61.9 & 45.2 \\
\quad + Self-Consistency & 65.7 & \textbf{85.5} & 74.3 & 49.1 & 84.0 & 62.0 & 53.5 & 90.1 & 67.2 & 43.0 & 76.0 & 54.9 & 36.0 & 66.5 & 46.8 \\
\rrnr  & 60.2 & 84.2 & 70.2 & 44.7 & \textbf{84.8} & 58.6 & 52.0 & \textbf{92.7} & 66.6 & 36.6 & \textbf{79.2} & 50.1 & 34.1 & \textbf{73.9} & 46.7 \\
\bottomrule
\end{tabular}
}
\caption{Comparison of relevance classification performance (Precision, Recall, and F1-score) across Qwen2.5-7B reranker variants.}
\label{tab:binary_classification_accuracy}
\end{table*}

Our results up to this point demonstrated that, under the exact same training regime,  rerankers that are trained to simply output a relevance prediction (\nrr) outperform rerankers trained to reason prior to making the relevance prediction (\rr), on average. But, \textit{what if we disable the reasoning for \rr?} We hypothesize that if the reasoning is crucial to \rr's relevance prediction, its reranking accuracy should drop if it does not reason.

\paragraph{Experiment Setup.} In order to disable the reasoning for \rr (\rrnr), we pre-fill the LLM's reasoning with a "forced" reasoning process: \texttt{<think> Okay, I think I have finished thinking. </think>}, following the setup from \citet{ma2025reasoning}.  We then follow the same evaluation setup as in Section \ref{sec: rq1}. Note that this, in essence, turns \rr into a standard pointwise reranker as it only needs to output the relevance label. 

\paragraph{Results.} The results of this experiment can be found in Table \ref{tab:no_reasoning}. For MS MARCO, \rrnr is consistently more effective than \rr, improving by an average of 0.8, 0.5, and 1.4 points across the 1.5B, 3B, and 7B model sizes, respectively. On BRIGHT, \rrnr is less effective for smaller models (1.5B and 3B), but as model size increases, \rrnr begins outperform \rr. In fact, at the 7B scale, \rrnr is able to improve by 3 points versus \rr on BRIGHT, closing the gap between \nrr and \rr.

These findings are remarkably concordant with those of Section \ref{sec: rq1}: (1) For in-domain datasets, reasoning reduces reranking effectiveness across all model sizes and (2) reasoning appears to be more beneficial out-of-domain for smaller rerankers, but as the LLM size increases, any benefits diminish, and reasoning actually \textit{hurts} reranking accuracy — even for \rr, which was trained to reason prior to making a relevance prediction.
\begin{figure*}[t!]
  \centering
  \includegraphics[width=\textwidth]{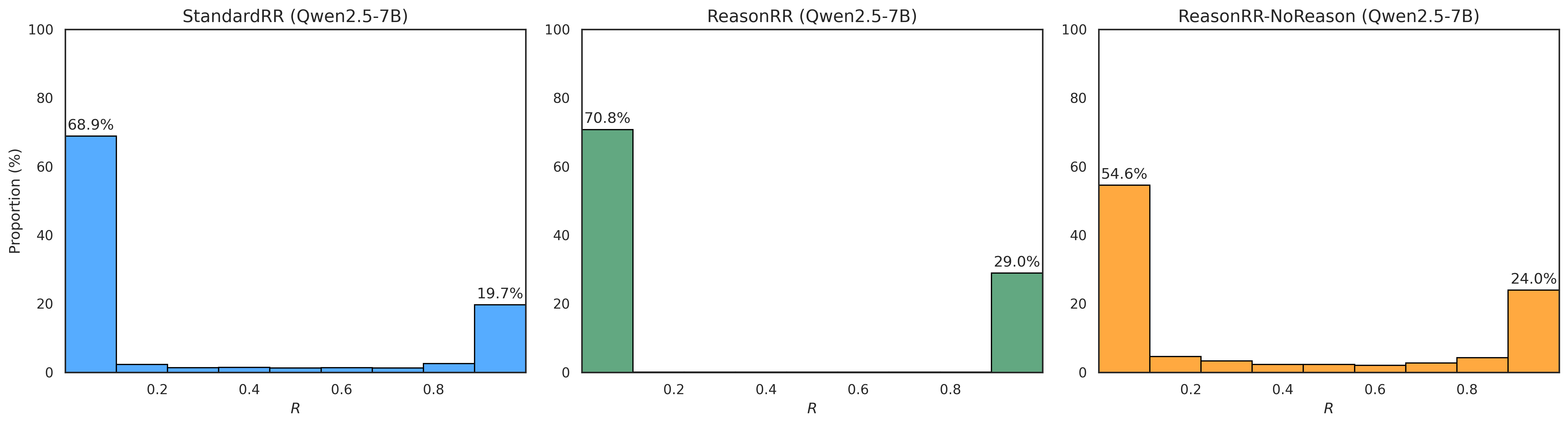}
  \caption{Relevance Scores Distribution across Qwen2.5-7B reranker variants on DL19. }
  \label{fig:logits}
\end{figure*}

\section{Why Does Reasoning Hurt Rerankers?}

One reason why \rr may perform worse than \nrr is that \rr has poorly calibrated and polarized probabilities for ranking due to the conclusions made by its reasoning process. For example, \rr will almost always assign very high probabilities when its reasoning concludes that a passage is relevant, and thus may not be able to reflect that one passage may be \textit{more} relevant than another passage. On the other hand, as \nrr is trained to only output ``true'' or ``false'', it may implicitly learn to output scores that account for one passage being more relevant than another passage. Due to this, we hypothesize that \nrr can better model the \textit{partial} relevance of query-passage pairs, making the outputs less polarized and preserving the uncertainty of scores which can be essential for the effectiveness of pointwise rerankers.

In this section, we dive deeper into this hypothesis. First, we investigate how \rr compares to \nrr and \rrnr as a simple binary relevance classifier. Then, we compare the relevance score distributions for \nrr, \rr, and \rrnr and examine a qualitative example of \rr's reasoning process. Finally, we discuss the results and propose potential improvements for \rr.

\subsection{Relevance Classification Comparison}
\label{sec:relevance_classification}
We first study how different reranking methods compare as simple relevance classifiers, ignoring their reranking accuracy measured by metrics like NDCG@10, which, ultimately, is what we \textit{really} care about. Doing so will allow us to better understand how much we can attribute differences in effectiveness to simply being worse relevance classifiers. In most cases, better relevance classification \emph{should} result in better reranking accuracy. 

For this experiment, we set $y_{pred} = 1$ if  $R$ $> 0.5$, and $y_{pred} = 0$ otherwise. For the ground truth relevance judgements, we set judgments > 2 (corresponding to highly relevant and perfectly relevant) as positive labels, and the rest as negative labels, following standard practice used for binary measures in IR \cite{macavaney2022adaptive}. The results for the Qwen2.5-7B models are in Table \ref{tab:binary_classification_accuracy}. 

 Comparing \nrr to \rr, we find that in terms of F1-score and precision, \nrr is consistently stronger than \rr. However, \rr generally has higher recall than  \nrr, indicating that \rr is classifying passages as relevant more frequently. We note that this is further confirmed in Figure \ref{fig:logits}, which we discuss in the next subsection. Surprisingly, \rrnr is generally worse at relevance classification than \rr (in terms of F1 and precision), yet outperforms it in terms of retrieval metrics, as discussed in Section \ref{sec:rq2}. Over the next two subsections, we provide potential explanations for this observation. 

\subsection{Relevance Scores Distribution}
\label{sec:score_distr}
Our observations in Section \ref{sec:relevance_classification} revealed a mismatch between relevance classification precision and reranking accuracy metrics (i.e., NDCG@10) for \rr versus \rrnr. To better understand why this may be the case, we plot the distribution of the relevance scores across the Qwen2.5-7B rerankers, shown in Figure \ref{fig:logits}. 

We find that \nrr and \rr place a similar proportion of their predictions in the low-relevance bin (0–0.1) for around 70\% of its scores. However, while \nrr spreads its remaining scores across both partial-relevance (0.1–0.9) regions (11.4\%) and high-relevance (0.9–1.0) regions (19.7\%), \rr exhibits a more extreme distribution. It places almost no scores in intermediate regions, assigning essentially any passage with partial relevance exclusively into the very high relevance bins (29.0\%). This demonstrates that the reasoning process leads the model to make polarized decisions, either relevant or not relevant, and rarely assigning partial relevance, matching our hypothesis. 

Interestingly, if we take \rr and forcefully remove its reasoning process as done in Section \ref{sec:rq2} (i.e., \rrnr), this no longer becomes the case. While the majority of the relevance scores are still at the tails of the distribution, \rr begins to place more of its predictions across the partial-relevance bins (21.4\%), while placing less predictions in the low-relevance and high-relevance bins. 

\subsection{Qualitative Study}
\label{sec:qualitative_study}

\begin{table*}[t!]
\centering
\small
\begin{tabular}{p{0.1\textwidth} p{0.8\textwidth}}
\toprule
\bf Query & how to help a jammed finger \\
\midrule
\bf Passage & Jammed finger vs. broken finger. Share on Pinterest. A broken finger is often more painful than a jammed finger, although both may be treated using a splint. A broken finger will cause severe pain and swelling that lasts for hours or even days. While a jammed finger is painful, it is not usually severe. \\
\midrule
\textbf{Reasoning} & Okay, let's see. The user is asking how to help a jammed finger. The passage provided talks about the difference between a jammed finger and a broken finger. It mentions that both might be treated with a splint, but a broken finger is more painful and has more swelling. First, the query is about treatment for a jammed finger. The passage does mention that a jammed finger can be treated with a splint, which is a direct answer to the query. \hl{However, the main focus of the passage is comparing jammed and broken fingers, not giving detailed treatment steps.} It does say that both might use a splint, which is relevant. But the passage doesn't go into other treatment methods like icing, elevation, or when to see a doctor. \hl{So, the passage is somewhat relevant because it mentions a treatment (splint) for a jammed finger.} But it's not comprehensive. The user might need more information, but the passage does provide a basic answer. However, the question is whether the passage is relevant, not if it's complete. \hl{Since it does address treatment (splint) for a jammed finger, even briefly, it's relevant.} The comparison to a broken finger might also help in understanding when a jammed finger isn't severe. \hl{Therefore, the answer is true.} \\
\midrule
\textbf{$R$} & 
\begin{tabular}[t]{@{}l@{}}
\textbf{\nrr}: 0.374 \\
\textbf{\rr}: 0.999 \\
\textbf{\rrnr}: 0.810 \\
\textbf{\rr + Self-Consistency}: 0.877
\end{tabular} \\
\bottomrule
\end{tabular}   
\caption{An example of \rr's (Qwen2.5-7B) reasoning process where it scores a partially relevant passage as highly relevant. As reference, we additionally provide the probability of relevance, $R$, for all reranker variants for the provided query-passage pair.}
\label{tab:qual}
\end{table*}

Finally, to better understand how \rr handles partially relevant passages, we examine the reasoning process of \rr for a random example in which \nrr produces a partially relevant prediction (i.e., in the 0.1-0.9 bin). The qualitative example can be found in Table \ref{tab:qual}. Looking at the reasoning process, we find that \rr explicitly mentions that the passage is "somewhat relevant", but because it has to select between two binary options (relevant or not relevant), it reasonably selects the passage as relevant. This subsequently makes $R$ high (0.999) as \rr finishes its reasoning stating "Therefore, the answer is true". This example shows how the reasoning process can bottleneck the ability for \rr to measure any concept of partial relevance,  even when the model states that the passage is not fully addressing the query.

\subsection{Discussion}

\begin{table}[t!]
\centering
\resizebox{\columnwidth}{!}{
\begin{tabular}{l|cc}
\toprule
& MS MARCO & BRIGHT \\
\midrule
\nrr  & 62.4  & 31.0  \\
\rrnr  &  58.8 &  30.8  \\
\midrule
\rr  &  57.4 & 27.8  \\
\quad + Self-Consistency &  \textbf{59.2} & \textbf{30.7} \\
\bottomrule
\end{tabular}}
\caption{Influence of Self-Consistency on \rr. For this experiment, the base LLM is Qwen2.5-7B. \textbf{Bold} results denote best between \rr and \rr + Self-Consistency. See Appendix \ref{sec:full_results_ablations} for results on individual datasets.}
\label{tab:self-consistency}
\end{table}
The findings from Sections \ref{sec:relevance_classification} and \ref{sec:score_distr} suggest that the ability to assign partial relevance is critical to the effectiveness of pointwise rerankers, not just the ability to predict if a passage is relevant or not. While \rr achieves higher effectiveness on binary relevance classification metrics compared to \rrnr, it still falls behind \rrnr in terms of NDCG@10.

This observation suggests that the improvements of \nrr over \rr arise not only from (1) \nrr being a better relevance classifier, but also from (2) \nrr having a stronger ability to capture partial relevance. As \rr and \rrnr only differ in their use of reasoning and \rrnr is more effective  in reranking, it would appear that (2) is the more crucial factor for reranking accuracy, highlighting the importance of partial relevance modeling for pointwise rerankers.

\begin{figure}[t]
\centering
\includegraphics[width=1.0\linewidth]{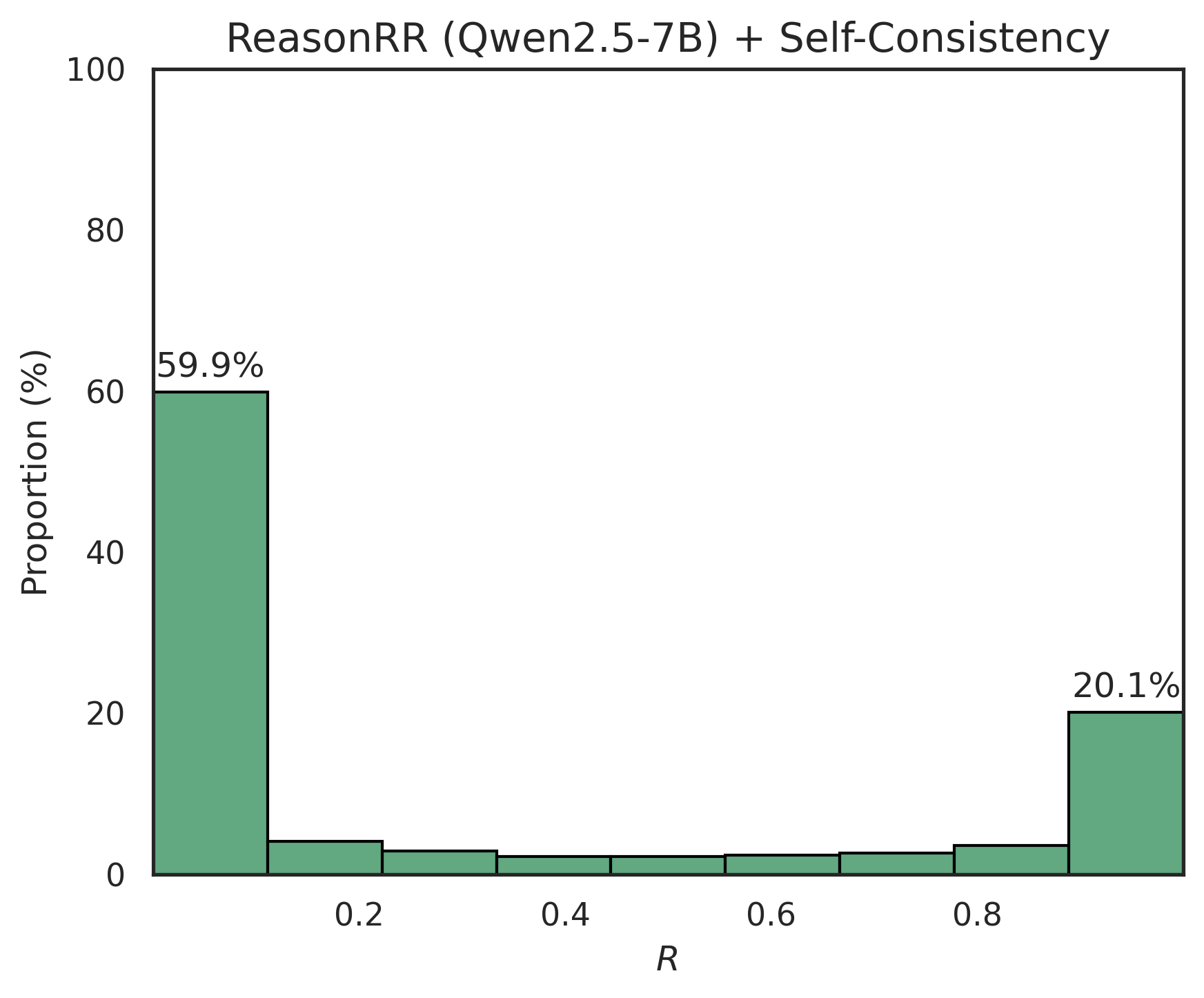}
\caption{Relevance Scores Distribution for \rr + Self-Consistency on DL19}
\label{fig:self-consistency}
\end{figure}

\paragraph{Can injecting a concept of partial relevance into \rr help?}  We explore a simple method to incorporate partial relevance into \rr: self-consistency \cite{wang2023selfconsistency}, which we denote as \rr + Self-Consistency. Unlike the majority vote approach used by \citet{wang2023selfconsistency}, we average the predicted $R$ values across eight sampled outputs from \rr to produce a continuous score suitable for reranking. The results of this experiment can be found in Table \ref{tab:self-consistency} and its relevance distribution is shown in Figure \ref{fig:self-consistency}. By leveraging self-consistency decoding, \rr begins to distribute its relevance scores away from the low-relevance (0–0.1) and high-relevance (0.9–1.0) bins and distributes~20\% of its predictions into partial-relevance (0.1–0.9) bins. By doing so, its NDCG@10 improves by 1.8 points on MS MARCO and 2.9 points on BRIGHT, even though the relevance classification metrics presented in Table \ref{tab:binary_classification_accuracy}, particularly precision, is generally on-par with \rr. 

\paragraph{So, is reasoning truly necessary for pointwise rerankers?} Even with the improvements from \rr + Self-Consistency, \rr still falls behind \nrr, suggesting that, at least in their current state, reasoning may not be best suited for pointwise reranking schemes. This is especially true when taking into account the lower inference costs of \nrr versus \rr.  

\paragraph{Potential Solutions.}
Given our results, we believe that to fully realize the benefits of reasoning in pointwise reranking, it is essential to re-design \rr to explicitly consider partial relevance, and the promising directions are:
\begin{itemize}
    \setlength\itemsep{0em}
    \item \emph{Training with non-binary relevance scores}: Instead of predicting binary relevance, \rr can be trained to generate graded scores (e.g., from 1 to 5). However, the current Rank1 training data~\citep{weller2025rank1} only provides binary labels, so it will be necessary to develop methods to synthesize realistic data that can accurately reflect partial relevance. We leave this to future work.
    \item \emph{Leveraging reasoning signals}: When \rr explicitly indicates partial relevance, through phrases like ``somewhat relevant'', these signals could be extracted to produce more accurate intermediate scores.
    \item \emph{Score calibration through loss function design}: Another approach is to directly train \rr to produce calibrated scores using tailored loss functions, encouraging outputs that reflect various degrees of relevance.
\end{itemize}

While each of these directions is worth further exploration, they remain open research problems to unlock the full potential of \rr. Until then, our results suggest that practitioners are better served by standard pointwise rerankers, which are simpler, more efficient, and currently more accurate.

\section{Related Work}

\paragraph{Reasoning and Retrieval.} Recent advances in LLMs have motivated exploration into how reasoning processes can be integrated into retrieval systems. O1 Embedder~\cite{yan2025o1} trained an LLM to generate intermediate "thoughts" based on the user query which were then used to enrich the query representations for dense retrieval. DEBATER~\cite{ji2025learning}, on the other hand, leveraged an iterative step-by-step reasoning process to learn more effective document representations.  Rank1~\cite{weller2025rank1} and Rank-R1~\cite{zhuang2025rank} instead focused on reasoning for rerankers, with Rank1 focusing on pointwise rerankers and Rank-R1 on Setwise ~\cite{zhuang2024setwise} rerankers, where the LLM reranker generates reasoning steps before selecting the most relevant document among a set of candidate documents. More recently, ReasonIR~\cite{shao2025reasonir}  explored the use of synthetic data to train retrievers for reasoning-intensive retrieval tasks.

\paragraph{Effcient Reasoning.} Another line of work parallel to ours has focused on making reasoning models more efficient, studying if the reasoning chain can be made more concise. Most closely related to our work is \citet{ma2025reasoning}, who demonstrated that the reasoning process of current reasoning models is not required for high performance. Another line of work, as summarized in \citet{sui2025stop}, has focused on fine-tuning LLMs to reason more efficiently by leveraging variable length chain-of-thought data \cite{munkhbat2025self, xia2025tokenskip, yu2024distilling, kang2025c3ot}. We highlight that rather than trying to make reasoning models more efficient, our work is primarily focused on questioning the necessity of reasoning for passage reranking, and \emph{not} making reasoning rerankers more efficient.
\section{Conclusion}

In this work, we study whether scaling test-time compute—via generation of reasoning tokens prior to making a relevance prediction—actually improves the accuracy of pointwise rerankers. To do so, we train and evaluate three pointwise rerankers, \nrr, \rr, and \rrnr. Through experiments across in-domain and out-of-domain datasets, we find that the reasoning process consistently \textit{harms} the accuracy of pointwise rerankers, especially as LLM size increases. Investigating the root cause of this result, we observe that the reasoning process restricts \rr's ability to capture partial relevance between query-document pairs, which is an important factor for pointwise reranking accuracy. While we explored self-consistency as a potential remedy for this restriction, \nrr  still outperformed \rr + Self-Consistency.

Our findings suggest that the reasoning process is unnecessary for pointwise rerankers and that practitioners are better served with simpler methods like \nrr. We believe that in order to fully realize the benefits of reasoning, it is essential to re-design how reasoning is utilized by \rr. Some promising directions we discuss include training with loss functions that encourage calibrated scoring or generating synthetic data that elicit relevance scores beyond binary labels. However, we emphasize that any improvements in the training of \rr should also be properly compared against strong and simple baselines.

\section*{Limitations}

\paragraph{Other Reranking Methods.} While we demonstrate that reasoning hurts pointwise rerankers, it remains an open question what influence reasoning may have for other reranking approaches such as listwise~\cite{sun2023chatgpt} and setwise~\cite{zhuang2024setwise}. However, as shown in~\citet{zhuang2024setwise}, pointwise rerankers are much more efficient as they can rerank candidate passages in parallel~\cite{ma2024fine} and thus, our study covers a very popular and commonly used approach for passage reranking.

\paragraph{LLM Models and Scales.} We limit our study to the Qwen2.5 family of models as it was the primary model used in related work~\cite{zhuang2025rank, weller2025rank1} and allowed us to control for factors such as LLM scale. However, as future work, it would be interesting to study the impact of reasoning across different model families. Additionally, our experiments were also limited to LLMs with $\leq$ 7B parameters. While our results showed that increasing LLM size benefited \nrr more than \rr, the influence at larger scales remains an open question.   We note that our \nrr at a 7B scale still outperforms or is competitive with the reported results for Rank1-14B and Rank1-32B on BRIGHT, which we believe can mitigate these concerns. 

\paragraph{Improvements to \rr.} Even though \rr + Self-Consistency — which was grounded in observations from our analysis — makes strong improvements on \rr, it still is less effective than our \nrr. While we propose potential solutions to improve \rr, we leave the implementation of these methods as future work. We hope our results and analysis can help in the development of new reasoning pointwise rerankers. 

\section*{Ethics Statement}

Our research solely uses publicly available datasets, and no personal information is collected. All datasets and models are used in accordance with its intended use and licenses. The goal of our study is to better understand the factors that influence the accuracy of LLM rerankers, which we hope can have a positive impact on building better search engines and other applications built on retrieval systems. 

While our results showed that standard pointwise rerankers, which minimize the output tokens generated by an LLM, outperform more verbose reasoning pointwise rerankers, we do recognize that such systems still rely on LLMs, which means that there is a risk that the LLM can produce biased, harmful, or offensive output. 

\section*{Acknowledgments}

DISTRIBUTION STATEMENT A. Approved for public release. Distribution is unlimited.
This material is based upon work supported by the MIT Comp Sci \& Artificial Intelligence L under Air Force Contract No. FA8702-15-D-0001 or FA8702-25-D-B002. Any opinions, findings, conclusions or recommendations expressed in this material are those of the author(s) and do not necessarily reflect the views of the MIT Comp Sci \& Artificial Intelligence L. \textcopyright 2025 Massachusetts Institute of Technology. Delivered to the U.S. Government with Unlimited Rights, as defined in DFARS Part 252.227-7013 or 7014 (Feb 2014). Notwithstanding any copyright notice, U.S. Government rights in this work are defined by DFARS 252.227-7013 or DFARS 252.227-7014 as detailed above. Use of this work other than as specifically authorized by the U.S. Government may violate any copyrights that exist in this work.

\bibliography{custom}

\clearpage
\appendix
\label{sec:appendix}
\section{Dataset Details}

We show the number of test queries for each dataset used for evaluation in Table \ref{tab:dataset_details}. 

\begin{table}[h!]
  \centering
  \begin{tabular}{l|c}
    \hline
    Dataset & \# Queries \\
    \hline
    TREC DL19 & 43 \\
    TREC DL20 & 54 \\
    TREC DL21 & 53 \\
    TREC DL22 & 76 \\
    TREC DL23 & 82 \\
    Biology & 103 \\
    Earth Science & 116 \\
    Economics & 103 \\
    Psychology & 101 \\
    Robotics & 101 \\
    Stackoverflow & 117 \\
    Sustainable Living & 108 \\
    Leetcode & 142 \\
    Pony & 112 \\
    AoPs & 111 \\
    TheoremQA Questions & 194 \\
    TheoremQA Theorems & 76 \\
    \hline
  \end{tabular} 
  \caption{Dataset Details}
  \label{tab:dataset_details}
\end{table}

The above datasets have the following licenses. 
\begin{itemize}
    \item The MS MARCO datasets are intended for non-commercial research purposes.
    \item BRIGHT is under CC BY 4.0 license.
    \item Rank1 training data, described in Section \ref{sec: rq1},  is under MIT License.
\end{itemize}

\begin{table}[t!]
\centering
\resizebox{\columnwidth}{!}{%
\begin{tabular}{l|cc|ccc}
\toprule
& \multicolumn{2}{c|}{MS MARCO v1} & \multicolumn{3}{c}{MS MARCO v2} \\
\cmidrule(r){2-3} \cmidrule(r){4-6}
& DL19 & DL20 & DL21 & DL22 & DL23 \\
\midrule
BM25 & 50.6 & 48.0 & 44.6 & 26.9 & 26.3 \\
\midrule
+ Qwen2.5-1.5B &&&&&\\
\quad \nrr & 73.1	 & 69.4  &	68.9  &	50.7	 & 44.2 \\
\quad \rr & 68.7	& 63.1	& 65.7	& 43.3	& 38.8 \\
\quad \rrnr & 69.9	& 61.7 &	67.3 &	44.9 &	39.5         \\
\midrule
+ Qwen2.5-3B &&&&&\\
\quad \nrr & 72.5	& 68.9	& 69.4	& 51.4	& 45.5 \\
\quad \rr& 70.4 & 66.4 & 65.9 & 45.2 & 41.3\\
\quad \rrnr  & 71.8 &	63.7 &	66.8 &	47.1 &	41.9         \\
\midrule
+ Qwen2.5-7B &&&&&\\
\quad \nrr & 74.6 & 70.0 & 70.9 & 50.3 & 46.3 \\
\quad \rr & 70.3 & 64.3 & 65.9 & 45.6 & 41.1\\
\quad \rrnr   & 73.3& 	65.0 &	69.1 &	46.1 &	40.5        \\
\quad \rr + Self-Consistency  & 71.5 &	66.7 &	68.8 &	46.0 &	42.9 \\
\bottomrule
\end{tabular}
}
\caption{Full results for \nrr, \rr, \rrnr, and \rr + Self-Consistency. All models rerank the top-100 passages from BM25.}
\label{tab:msmarco_appendix}
\end{table}

\begin{table*}[t]
\centering
\resizebox{\textwidth}{!}{
\begin{tabular}{l|rrrrrrr|rr|rrr|r}
\toprule
& \multicolumn{7}{c|}{StackExchange} & \multicolumn{2}{c|}{Coding} & \multicolumn{3}{c|}{Theorem-based} & \multirow{2}{*}{Avg.} \\
\cmidrule(r){2-8} \cmidrule(r){9-10} \cmidrule(r){11-13}
& Bio. & Earth. & Econ. & Psy. & Rob. & Stack. & Sus. & Leet. & Pony & AoPS & TheoQ. & TheoT. & \\
\midrule
BM25 + GPT-4 CoT              & 53.6 &	54.1 &	24.3 &	38.7 &	18.9 &	27.7 &	26.3 &	19.3 &	17.6 &	3.9 &	19.2 &	20.8 &	27.0 \\
\midrule
+ Qwen2.5-1.5B  &&&&&&&&&&&&&    \\
\quad \nrr   & 37.0 &	21.7 &	16.8 &	23.1 &	16.1 &	10.0 &	26.3 &	2.6 &	30.6 &	1.8 &	16.1 &	26.1 &	19.0	 \\ 
\quad \rr     & 32.5	& 20.3 & 	12.3 & 	25.5 & 	11.1 & 	15.3 & 	23.5 & 	6.6 & 	12.3 & 	3.4 & 	10.6 & 	13.7 &	15.6	 \\
\quad \rrnr  & 23.1	& 15.3 &	10.4 &	13.3 &	10.4 &	6.2 &	7.0 &	4.4 &	11.3 &	3.1 &	12.0 &	23.2 &	11.6          \\
\midrule
+ Qwen2.5-3B  &&&&&&&&&&&&&   \\
\quad \nrr    & 41.6 &	27.1 &	  20.9 &	31.9 &	22.2 &	16.9 &	30.3 &	13.2 &	42.0 &	2.7 &	16.2 &	30.6 &	24.6	 \\
\quad \rr     &  37.3 &	27.8 &	20.7 &	33.1 &	18.3 &	24.3 &	25.2 &	11.3 &	26.2 &	4.7 &	20.7 &	34.0 &	23.6\\
\quad \rrnr   & 40.8  &	20.5  &	20.3  &	31.9  &	14.0  &	15.3  &	23.3  &	18.7  &	37.3  &	3.7  &	24.6  &	31.1 & 23.4         \\
\midrule
+ Qwen2.5-7B  &&&&&&&&&&&&&   \\
\quad \nrr          &  47.1 &	38.0 &	28.1 &	44.1 &	26.1 &	29.5 &	36.5 &	19.3 &	37.5 &	4.6 &	22.4 &	39.4 &	31.0	 \\ 
\quad \rr            & 47.0 &	35.4 &	24.0 &	35.2 &	20.0 &	25.2 &	31.0 & 	15.1&	36.0 &	5.9	 & 22.2	& 36.6 &	27.8\\
\quad \rrnr    & 56.0  &	41.9  &	27.5  &	38.5  &	23.2  &	21.6  &	32.7  &	16.3  &	39.4  &	7.2  &	27.2  &	38.0  &	30.8       \\
\quad \rr + Self-Consistency  & 49.6 &	38.2 &	27.4 &	40.9 &	23.7 &	29.3 &	33.2 &	14.9 &	38.4 &	8.1 &	25.4 &	39.1 &	30.7\\
\midrule
+ Rank1-7B (Our Results) & 48.0 & 37.2 & 21.8 & 35.1 & 19.9 & 22.6 & 31.0 & 12.7 & 30.8 & 6.8 & 26.0 & 38.2 & 27.5 \\
+ Rank1-7B (Reported Results) & 48.8 & 36.7 & 20.8 & 35.0 & 22.0 & 18.7 & 36.2 & 12.7 & 31.2 & 6.3 & 23.7 & 37.8 & 27.5 \\
\bottomrule
\end{tabular}
}
\caption{Full results for \nrr, \rr, \rrnr, and \rr + Self-Consistency. All models rerank the top-100 passages from BM25 + GPT-4 CoT .}
\label{tab:bright_appendix}
\end{table*}

\section{Model Details}

\begin{itemize}
    \item $\texttt{Qwen2.5-1.5B}$: A 1.5B base model. Huggingface ID: $\texttt{Qwen/Qwen2.5-1.5B}$ 
    \item $\texttt{Qwen2.5-3B}$: A 3B base model. Huggingface ID: $\texttt{Qwen/Qwen2.5-3B}$ 
    \item $\texttt{Qwen2.5-7B}$: A 7B base model. Huggingface ID: $\texttt{Qwen/Qwen2.5-7B}$

\end{itemize}

The above models have the following licenses. 
\begin{itemize}
    \item $\texttt{Qwen2.5-1.5B}$ is under the Apache 2.0 License. 
    \item $\texttt{Qwen2.5-3B}$ is under the Qwen Research License Agreement.
    \item $\texttt{Qwen2.5-7B}$ is under the Apache 2.0 License. 
\end{itemize}
We also leverage Pyserini \cite{lin2021pyserini} and vLLM~\cite{kwon2023efficient} which are under the Apache 2.0 License.

\section{Training and Inference Details for \nrr and \rr}

To train \nrr and \rr  we fine-tune Qwen2.5 using LoRA \cite{hu2022lora} for one epoch with rank 32 and alpha 64, using a batch size of 128 and a learning rate of 2e-4. We apply LoRA to all the linear layers of the transformer model. 
Note, to train the \nrr we leverage the same dataset as \rr, but only use the (query, passage, relevance label) triples, ignoring the R1 reasoning process. Training for each reranker took less than a day and was done on an A100 GPU. Due to limited computational resources, each model is only trained once. 

For inference, we run all models on NVIDIA A6000 (48GB) and A100 (80GB) GPUs. As the \nrr and \rr outputs are run with greedy decoding, all the scores in the paper are from a single run. 

\section{Full Results for \rrnr and \rr + Self-Consistency}
\label{sec:full_results_ablations}

\begin{table}[t!]
\centering
\resizebox{\columnwidth}{!}{
\begin{tabular}{l|cc}
\toprule
& MS MARCO & BRIGHT \\
\midrule
\nrr  & 62.4  & 31.0  \\
\rrnr  &  58.8 &  30.8  \\
\midrule
\rr  &  57.4 & 27.8  \\
\quad + Self-Consistency (3 samples) &  59.1 & 30.6 \\
\quad + Self-Consistency (8 samples) &  59.2 & 30.7 \\
\bottomrule
\end{tabular}}
\caption{Influence of the number of sampled chains for \rr + Self-Consistency.}
\label{tab:self-consistency_ablation}
\end{table}

In this section, we provide the full results for \rrnr and \rr + Self-Consistency across MS MARCO and BRIGHT datasets. These results can be found Table \ref{tab:msmarco_appendix} and Table \ref{tab:bright_appendix}. 

\section{Comparison with Rank1~\cite{weller2025rank1}}
\label{sec:reproduction_rank1}

We also provide the results for Rank1 to ensure that our reproduction, \rr, is valid. These results are in Table \ref{tab:bright_appendix}. We found that the Rank1 paper used an earlier edition of BRIGHT, which had minor differences in queries and judged documents.  Thus, we report Rank1 results on the new BRIGHT, Rank1 (Our Results) as well as the original papers reported results, Rank1 (Reported Results). Comparing \rr for Qwen2.5-7B to Rank1-7B, both are comparable in terms of NDCG@10 (27.8 versus 27.5). 

\section{Number of Sampled Outputs for \rr + Self-Consistency}
In this section, we present the results of \rr  + Self-Consistency when we only sample 3 reasoning chains from \rr. The results are shown in Table \ref{tab:self-consistency_ablation}. We find that  \rr  + Self-Consistency (n=3) is as effective as \rr  + Self-Consistency (n=8), suggesting that sampling more reasoning chains is not more effective for \rr. 

\section{Prompts}

For training and evaluation of \nrr, \rr, \rrnr, and \rr + Self-Consistency, we leverage the same exact prompts used in the Rank1~\cite{weller2025rank1} paper, but apply the Qwen2.5~\cite{yang2024qwen2} chat template. Below we repeat the baseline prompt. For the dataset specific prompts we used, please refer to~\citet{weller2025rank1}. 

\bigskip
\noindent
\textbf{\nrr Standard Prompt:}
\begin{tcolorbox}
\texttt{<|im\_start|>}system \\
Determine if the following passage is relevant to the query. Answer only with 'true' or 'false'. \\
\texttt{<|im\_end|>} \\
\texttt{<|im\_start|>}user \\
Query: \{\} \\
Passage: \{\} \\
\texttt{<|im\_end|>} \\
\texttt{<|im\_start|>}assistant
\end{tcolorbox}
\newpage
\noindent
\textbf{\rr Standard Prompt:}

\begin{tcolorbox}
\texttt{<|im\_start|>}system \\
Determine if the following passage is relevant to the query. Answer only with 'true' or 'false'. \\
\texttt{<|im\_end|>} \\
\texttt{<|im\_start|>}user \\
Query: \{\} \\
Passage: \{\} \\
\texttt{<|im\_end|>} \\
\texttt{<|im\_start|>}assistant \\
\texttt{<think>} 
\end{tcolorbox}

\noindent
\textbf{\rrnr Standard Prompt:}

\begin{tcolorbox}
\texttt{<|im\_start|>}system \\
Determine if the following passage is relevant to the query. Answer only with 'true' or 'false'. \\
\texttt{<|im\_end|>} \\
\texttt{<|im\_start|>}user \\
Query: \{\} \\
Passage: \{\} \\
\texttt{<|im\_end|>} \\
\texttt{<|im\_start|>}assistant\\
\texttt{<think>}\\
Okay, I have finished thinking. \\
\texttt{</think>}
\end{tcolorbox}

\end{document}